\documentclass{sig-alternate}

\usepackage{subfigure}
\usepackage{graphicx}
\usepackage{url}

\makeatletter
\let\@copyrightspace\relax
\makeatother

\begin{document}

\title{Evaluating the SiteStory Transactional Web Archive With the ApacheBench Tool}

\numberofauthors{2}

\author{
\alignauthor
Justin F. Brunelle\\
       \affaddr{Old Dominion University}\\
       \affaddr{Department of Computer Science}\\
       \affaddr{Norfolk, Virginia, 23508}\\
       \email{jbrunelle@cs.odu.edu}
\alignauthor
Michael L. Nelson\\
       \affaddr{Old Dominion University}\\
       \affaddr{Department of Computer Science}\\
       \affaddr{Norfolk, Virginia, 23508}\\
       \email{mln@cs.odu.edu}
}

\maketitle
\begin{abstract}
\sloppy Conventional Web archives are created by periodically crawling a Web
site and archiving the responses from the Web server.  Although easy to
implement and commonly deployed, this form of archiving typically misses
updates and  may not be suitable for all preservation scenarios, for
example a site that is required (perhaps for records compliance) to keep
a copy of all pages it has served.  In contrast, transactional archives
work in conjunction with a Web server to record all content that has
been served.  Los Alamos National Laboratory has developed SiteStory, an
open-source transactional archive written in Java that
runs on Apache Web servers, provides a Memento compatible access
interface, and WARC file export features. We used Apache's ApacheBench
utility on a pre-release version of SiteStory to measure response time and
content delivery time in different environments and on different
machines. The performance tests were designed to determine the
feasibility of SiteStory as a production-level solution for high
fidelity automatic Web archiving.  We found that SiteStory does not
significantly affect content server performance when it is performing
transactional archiving.  Content server performance slows from 0.076
seconds to 0.086 seconds per Web page access when the content server is
under load, and from 0.15 seconds to 0.21 seconds when the resource has
many embedded and changing resources.\fussy
\end{abstract}





\section{Introduction}
\label{introduction}
Web archiving is an important aspect of cultural, historical, governmental, and even institutional memory. The cost of capturing Web-native content for storage and archiving varies and is dependent upon several factors. The cost of human-harvested Web archiving has prompted research into automated methods of digital resource capture. The traditional and classic method of automatic capture is the Web crawler, but recent migrations toward more personalized and dynamic resources have rendered crawlers ineffective at high-fidelity capture in certain situations. For example, a crawler cannot capture every representation of a resource that is customized for each user. Transactional archiving can, in some instances, provide an automatic archiving solution to this problem where crawlers fall short.

\subsection{Transactional Archiving}
\label{ta}

The purpose of a transactional archive (TA) is to archive every
representation of a resource that a Web server disseminates.  A client
does an HTTP GET on a URI and the Web server returns the representation
of the resource at that time.  At dissemination time, it is the
responsibility of TA software to send the representation seen by
the client to an archive.  In this way, \emph{all}
representations returned by the Web server can be archived.  If storing
all served representations is costly (e.g., a high-traffic site with
slowly changing resources), it is possible to optimize a TA in a
variety of ways: store only unique representations, store every
n$^{th}$ representation, etc.

\begin{figure*}[t]
\begin{center}
\includegraphics[width=140mm]{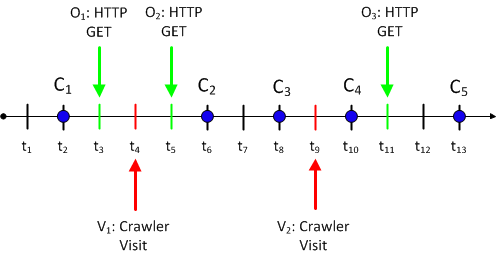}
\caption{User and crawler accesses control the archival interval, capturing each returned representation.}
\label{timeline}
\end{center}
\end{figure*}

Figure \ref{timeline} provides a visual representation of a typical page change and user access scenario. This scenario assumes an arbitrary page that will be called \emph{P} changes at inconsistent intervals. This timeline shows page \emph{P} changes at points $C_{1}$, $C_{2}$, $C_{3}$, $C_{4}$, and $C_{5}$ at times $t_{2}$, $t_{6}$, $t_{8}$, $t_{10}$, and $t_{13}$, respectively. A user makes a request for \emph{P} at points $O_{1}$, $O_{2}$, and $O_{3}$ at times $t_{3}$, $t_{5}$, and $t_{11}$, respectively. A Web crawler (that captures representations for storage in a Web archive) visits \emph{P} at points $V_{1}$ and $V_{2}$ at times $t_{4}$ and $t_{9}$, respectively. 

Since $O_{1}$ occurs after change $C_{1}$, an archived copy of $C_{1}$
is made by the TA. When $O_{2}$ is made, \emph{P} has not changed since
$O_{1}$ and therefore, an archived copy is not made since one already
exists. The Web crawler visits $V_{1}$ captures $C_{1}$, and makes a
copy in the Web archive.  In servicing $V_{1}$, an unoptimized TA will store
another copy of $C_{1}$ at $t_{4}$ and an optimized TA could detect that no 
change has occurred and not store another copy of $C_{1}$.

Change $C_{2}$ occurs at time $t_{6}$, and $C_{3}$ occurs at time
$t_{8}$. There was no access to \emph{P} between $t_{6}$ and $t_{8}$,
which means $C_{2}$ is lost -- an archived copy exists in neither the
TA nor the Web crawler's archive. However, the argument can be made
that if no entity observed the change, should it be archived?
Change $C_{3}$ occurs and the representation of \emph{P} is archived during the crawler's visit
$V_{2}$, and the TA will also archive $C_{3}$.  After $C_{4}$, a user
accessed \emph{P} at $O_{3}$ creating an archived copy of $C_{4}$ in
the TA.  

In the scenario depicted in Figure \ref{timeline}, the TA will have
changes {$C_{1}$, $C_{3}$, $C_{4}$}, and a conventional archive will
only have {$C_{1}$, $C_{3}$}.  Change $C_{2}$ was never served to any
client (human or crawler) and is thus not archived by either system.
Change $C_{5}$ will be captured by the TA when \emph{P} is accessed
next.


\subsection{SiteStory}
\begin{figure*}[htp!]
\begin{center}
\includegraphics[width=110mm]{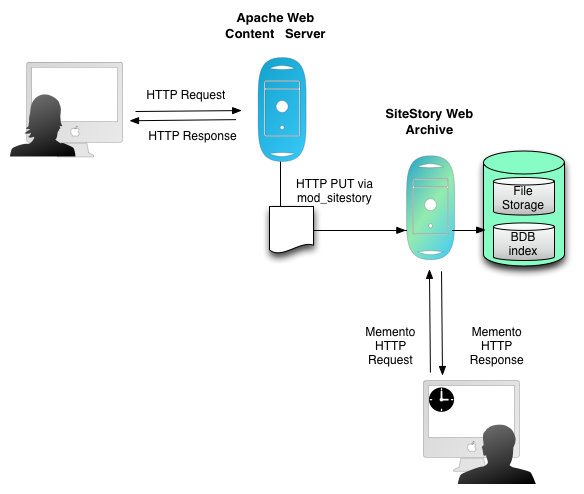}
\caption{SiteStory consists of two parts: mod\_sitestory which is installed on the Apache server to be archived, and the transactional archive itself.  Image taken from the SiteStory GitHub at \texttt{http://mementoWeb.github.com/SiteStory/}.}
\label{twa}
\end{center}
\end{figure*}

\sloppy Los Alamos National Laboratory has developed SiteStory\footnote{\url{http://mementoWeb.github.com/SiteStory/}}, an open-source transactional Web archive.  Figure \ref{twa} illustrates the components and process of SiteStory.  First, mod\_sitestory is installed on the Apache server that contains the content to be archived.  When the Apache server builds the response for the requesting client, mod\_sitestory sends a copy of the response to the SiteStory Web archive, which is deployed as a separate entity.  This Web archive then provides Memento-based access (see Section \ref{priorwork}) to the content served by the Apache server with mod\_sitestory installed, and the SiteStory Web archive is discoverable from the Apache Web server using standard Memento conventions (see Section 4 of \cite{memento:rfc}).

Sending a copy of the HTTP response to the archive is an additional task for the Apache Web server, and this task must not come at too great a performance penalty to the Web server.  The goal of this study is to quantify the additional load mod\_sitestory places on the Apache Web server to be archived.\fussy

\subsection{Organization and Purpose}
\label{organization}
This Technical Report details the work performed with SiteStory, and the results of the performance tests and benchmarking performed as part of a feasibility study. 
The rest if this Technical Report is organized as follows: Section \ref{priorwork} discusses the contributions of prior research efforts. Section \ref{experiment} discusses the experiment design and execution. Section \ref{results} details the results and findings of the experiment. Finally, Section \ref{conclusion} summarizes the findings and impacts of this Technical Report, and outlines the upcoming extensions of this work. 

\section{Prior Work}
\label{priorwork}
\begin{figure}[ht!]
\begin{center}
\includegraphics[width=65mm]{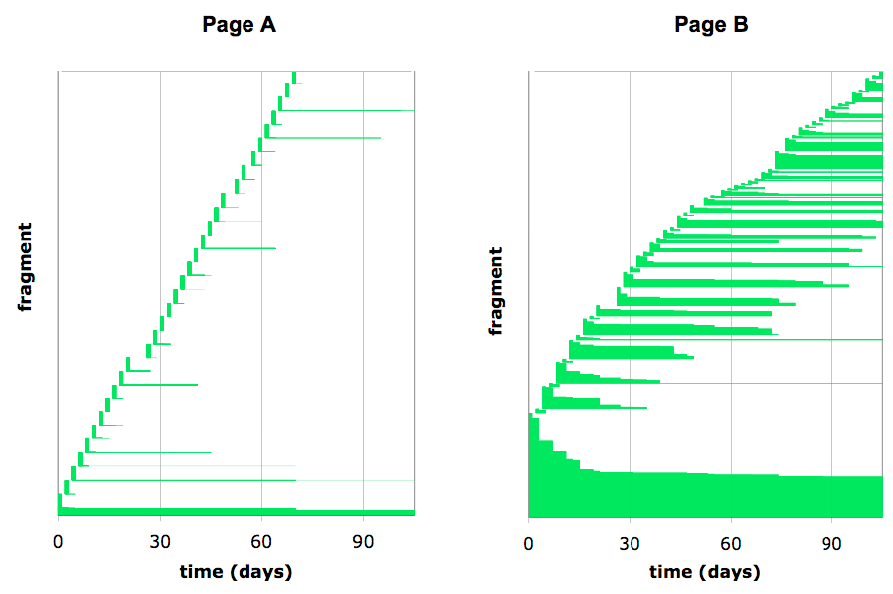}
\caption{ Page A shows rapidly appearing and disappearing content, while Page B shows longer-lived content. This image was originally published in Olston's 2008 paper \cite{olston2008recrawl}.}
\label{olston}
\end{center}
\end{figure}

Extensive research has been done to determine how Web documents change on the Web.  Studies of ``wild'' pages (such as Cho's work with crawlers \cite{cho2000evolution} or Olston's work in recrawl scheduling \cite{olston2008recrawl}) have shown that pages change extremely frequently.  Figure \ref{olston} (taken from Olston's paper) visually shows the ephemeral nature of information contained within a Web page.  In this figure, one can see that not only do pages change very frequently, but one can see that pages change in different ways.  In this figure, Page A has small sections of content that change rapidly.  This behavior is called ``churn''.  Page B has longer-lived content, but additional content is added to the page over time.  This is called ``scroll'' behavior.

Prior research has focused on crawlers and robots to find pages and monitor their change patterns \cite{brewington2000keeping, fetterly2004large, 511465}.  These crawlers follow the links on pages to discover other pages and archive and recrawl the discovered pages over time to compile an archive.  This method is unsuitable for an intranet that is closed to the public Web; crawlers cannot access the resources of archival interest \cite{hiddenurls}.  As a way to have finer control over the archival granularity, transactional archiving should be used. Transactional archiving implementations include TTApache \cite{Dyreson:2004:MVW:988672.988730} and pageVault \cite{transarch}.  TTApache is a server-side solution and pageVault operates on the client-side.  For each user access of a Web resource, TTApache compares a hash of the content and stores a copy at the server if it has changed, and pageVault determines if the content has changed by rendering the content on the client and archiving it locally if needed.  These implementations were also shown not to substantially increase the access time seen by Web users; pageVault saw an increase of access time from 1.1 ms to 1.5 ms, and TTApache saw a 5-25\% increase in response time, depending on requested document size.

Memento is a joint project between Old Dominion University and Los Alamos National Laboratory.  The Memento Framework defines HTTP extensions that allow content negotiation in the dimension of time \cite{nelson:memento:tr, ldow2010:memento}.  When used with Memento-aware user agents like MementoFox \cite{sandersonimplementing}, users can set a desired datetime in the past and browse the Web as it existed at (or near) that datetime.  Unlike other, single-archive applications like DiffIE \cite{Teevan:2010:LSH:1753326.1753530, 1622221}, Past Web Browser \cite{jatowt2006journey}, or Zoetrope \cite{adar2008zoetrope}, Memento provides an multi-archive approach to presenting the past Web. Integrating multiple Web archives can give a more complete picture of the past Web \cite{hmotwia}.

\section{Experiment design}
\label{experiment}

SiteStory was tested with a variety of loads, a variety of resources, and on two machines with different configurations and specifications. Three different tests were run during the experiment. The details of the experiment setup and execution is included in this section.

\subsection{Experiment Machines}
\label{machines}

The SiteStory benchmarking experiment was conducted with a pre-release
version of SiteStory installed on two machines, PC1 and PC2. Both
machines ran the prefork version of the Apache 2 Web server, and in
both cases the mod\_sitestory-enabled Apache server provided content
from \url{localhost:8080} and the SiteStory archive was installed at
\url{localhost:9999}.  Even though we installed SiteStory on different
ports on the same machine, it can be installed on two different
machines.  Although the developers have experimented with optimizations
discussed in Section \ref{ta}, SiteStory currently archives all
returned representations regardless of whether the representation has
changed or not.

PC1 has a single core 2.8 GHz processor.  PC1 has no memory remaining
on the server; it is 100\% utilized. PC1 represents a worst-case
scenario for a server -- it has been completely bogged down with
background processes. To simulate this load, a script runs throughout
the experiment that initiates requests for Web pages to create the load
on the server. PC2 has two 1GHz processors and is unhindered during the
testing by additional requests.  Both of these machines run Linux
Ubuntu; PC1 ran version 11, while PC2 ran version 10. These machines
complement each other by providing two extremes of a potential content
server: an overtaxed, under performing server and a high performing,
unburdened server. The results from each of these machines are provided
in Section \ref{results}.\\

\subsection{Experiment Runs}
\label{runs}
Three separate experiments were run, and each experiment was run on both machines PC1 an PC2. The first experiment tests the throughput of a content server enabled with SiteStory software. This experiment ran for 45 days, from January 14th, 2012 to February 28th, 2012. This is described in Section \ref{ab}. The second experiment performs a series of accesses to 100 static resources to test the access rates, response times, and round trip times possible. This test was run from March 1st, 2012 until March 16th, 2012. This is described in Section \ref{static}.  The third experiment performs a series of accesses to 100 dynamic, constantly changing set of 100 resources to demonstrate a worst-case scenario for SiteStory -- everything is archived on each access. This test was also run from March 1st, 2012 until March 16th, 2012. This final experiment is described in Section \ref{changes}. 

\subsection{Connection Handling: ab}
\label{ab}
This first experiment to measure the differences in throughput when SiteStory is running and when SiteStory is turned off was run twice a day (at 0700 and 1900 EST) for 45 days, resulting in 90 data points. The experiment utilized the ab (ApacheBench) tool\footnote{\texttt{http://httpd.apache.org/docs/2.0/programs/ab.html}}. This utility makes N number of connections as quickly as possible with C concurrency, where N and C are variables specified by the user. The ab utility records the response, throughput, and other server stats during a test. Essentially, the ApacheBench utility issues HTTP GET requests for content as quickly as possible to establish a benchmark for performance. A run in ab provides output similar to that seen in Appendix \ref{abApp}.

Three different HTML resources were targeted with this test: a small, medium, and large file of sizes 1kB, 250 kB, and 700 kB. These resource sizes were chosen after a brief survey of the average file sizes in a corporate intranet provided a minimum, average, and maximum file size of a Web page\footnote{These file sizes were empirically determined in an internal MITRE study.}. Four different connections (1,000, 10,000, 100,000, 216,000) and four concurrencies (1, 100, 200, 450) were used. The connection numbers of 1,000, 10,000, and 100,000 were chosen arbitrarily. The connection number of 216,000 was chosen after observing this as 2011's yearly maximum hourly rate of Website access\footnote{This access rate was chosen after observing the average 2011 Website access rate within MITRE's corporate intranet.}. Similarly, the access concurrencies of 1, 100, and 200 were chosen arbitrarily, but the concurrency of 450 was chosen as the observed average expected number of concurrent accesses to a site\footnote{This access concurrency was chosen after observing the average 2011 Website access rate within MITRE's corporate intranet.}. For simplicity and brevity, this work will only consider a subset of these test runs. This report discusses the runs of 10,000 connections with concurrencies 1 and 100, and runs of 216,000 connections with concurrencies 1 and 100. This subset of results illustrates typical results of all other tests. 

The three resources were modified between each set of connections to ensure the resource is archived each run. To modify the resources, a script was run to update the past run time of the script on each page, and change the image that was embedded in the page. These modifications would ensure that not only the image was changed and able to be re-archived, but the surrounding HTML was changed, as well. Since SiteStory re-archives content whenever a change is detected, each test run results in each resource being re-archived. It is essential to make sure the resource is re-archived to observe the effect of an archival action on the content server performance.

\begin{figure*}[htp]
  \begin{center}
    \subfigure[Total run time for the ab test with 10,000 connections and 1 concurrency.]{\label{tr1}\includegraphics[width=150mm]{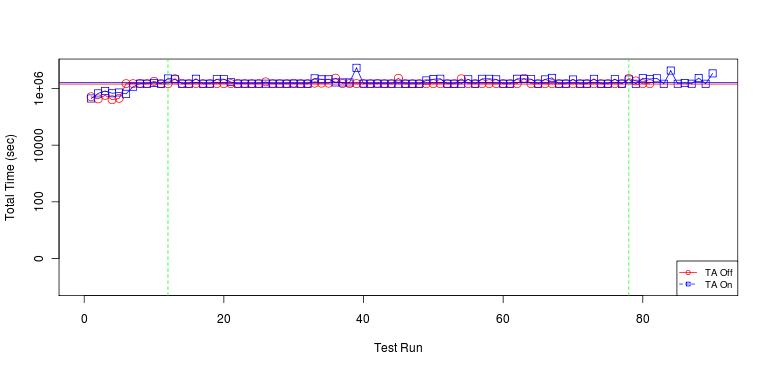}} \\
    \subfigure[Total run time for the ab test with 10,000 connections and 100 concurrency.]{\label{tr2}\includegraphics[width=150mm]{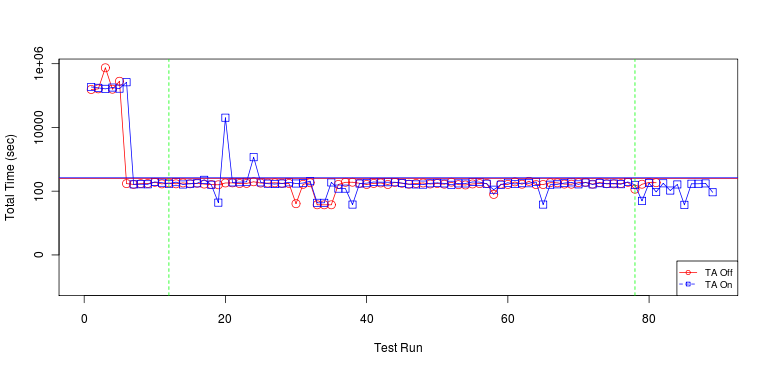}} 
  \end{center}
  \caption{Total run time for 10,000 Connections.}
  \label{tares}
\end{figure*}

Each ab test was performed twice: once while SiteStory was turned on, and once while it was turned off. This shows how SiteStory affects the content server performance. A subset of the results are provided in Figure \ref{tares}. The red lines represent the runs in which SiteStory was turned off, while the blue lines represent the runs in which SiteStory was turned on. Each entry on the x-axis represents an independent test run. The y-axis provides the amount of time it took to execute the entire ab run. The horizontal lines represent the averages over the entire experiment. The dotted, vertical green lines indicate machine restart times due to power outages. The power outages were noted to show when a cache and memory resets may have occurred that could impact the performance of the machines.
 
\begin{figure*}[htp]
  \begin{center}
    \subfigure[Total run time for the ab test with 216,000 connections and 1 concurrency.]{\label{tr3}\includegraphics[width=150mm]{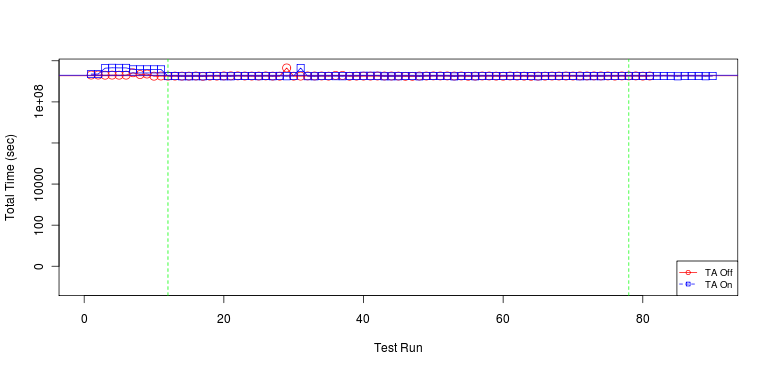}} \\
    \subfigure[Total run time for the ab test with 216,000 connections and 100 concurrency.]{\label{tr4}\includegraphics[width=150mm]{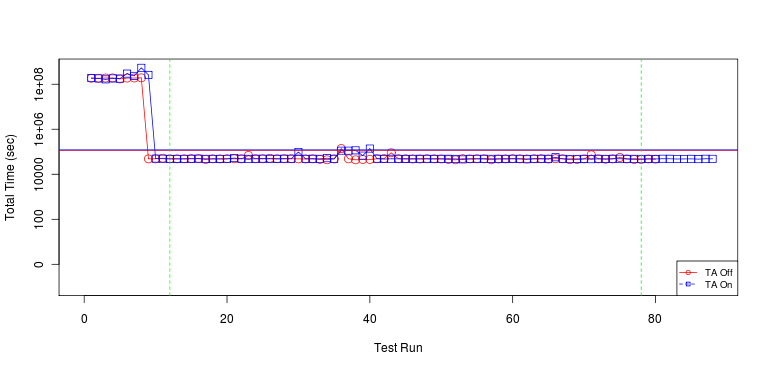}} 
  \end{center}
  \caption{Total run time for 216,000 Connections.}
  \label{tares2}
\end{figure*}

To illustrate how SiteStory affects the content server's performance, please reference Figures \ref{tares} and \ref{tares2} that portray the changes in the total run time of the ab test when SiteStory is on (actively archiving served content) and off (not archiving served content).

\subsection{100 Static Resources: Clearing the Cache}
\label{static}
The second experiment uses the \texttt{curl} command to access 100 different HTML resources, none of which change. After running the ab tests in Section \ref{ab}, a theory was formulated that a reason for some of the anomalies was from server caching.  This additional test shows the effect of clearing the server cache on SiteStory by accessing a large number of large files in sequence. This access essentially thrashes the server cache. Each resource has text, and between 0 and 99 images (the 0th resource has 0 images, the 1st resource has 1 image, etc.). These resources were generated by a Perl script that constructed 100 different HTML pages and embedded between 0-99 different images in the generated resources. The resources were created with different sizes, and different numbers of embedded resources to demonstrate how SiteStory affects content server performance with a variety of page sizes and embedded images.

\begin{figure*}[htp]
  \begin{center}
    \subfigure[Total access time for the 100 static resources on PC1.]{\label{ts1}\includegraphics[width=150mm]{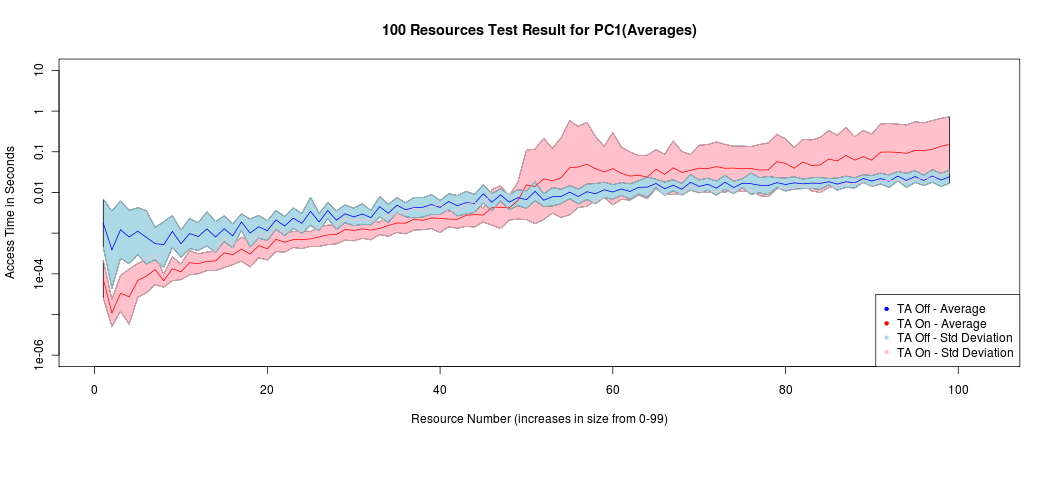}} \\
    \subfigure[Total access time for the 100 changing resources on PC1.]{\label{tc1}\includegraphics[width=150mm]{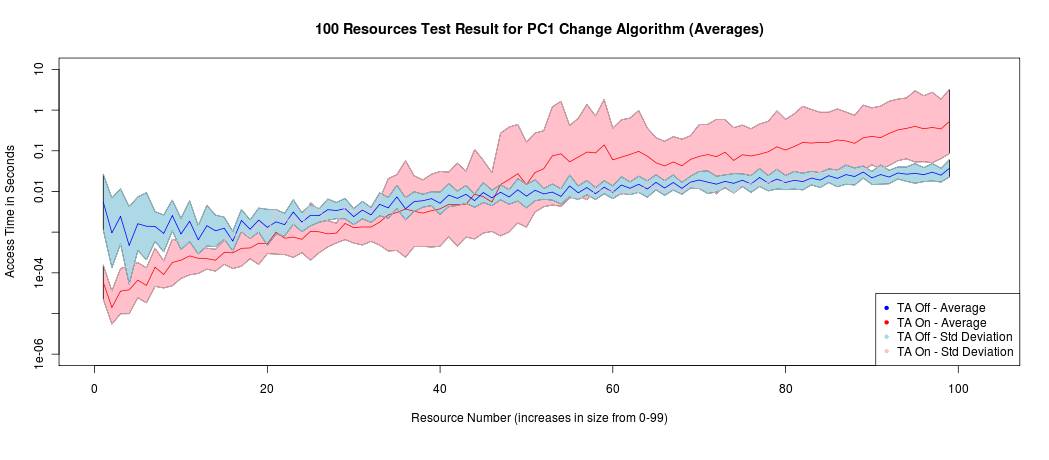}} 
  \end{center}
  \caption{100 resources accessed on PC1. Resource \emph{n} has \emph{n} embedded images.
}
  \label{pc100}
\end{figure*}

\begin{figure*}[htp]
  \begin{center}
    \subfigure[Total access time for the 100 static resources on PC2.]{\label{ts2}\includegraphics[width=150mm]{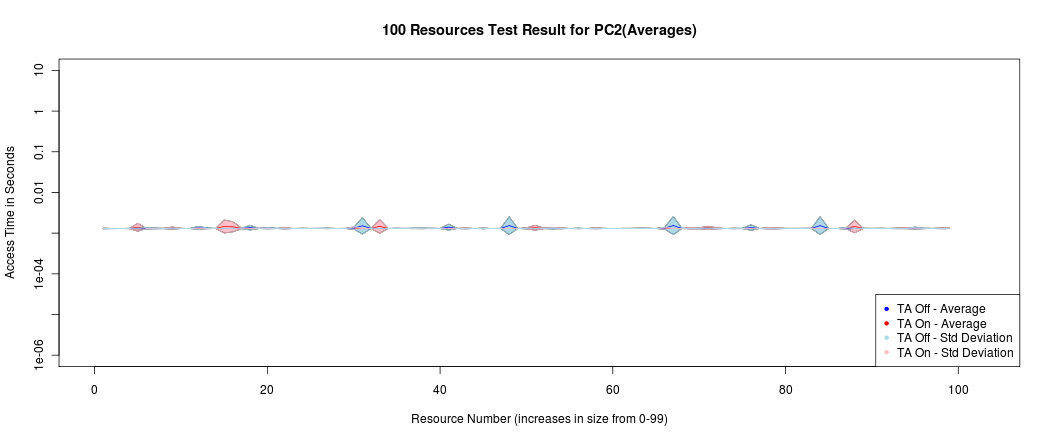}} \\
    \subfigure[Total access time for the 100 changing resources on PC2.]{\label{tc2}\includegraphics[width=150mm]{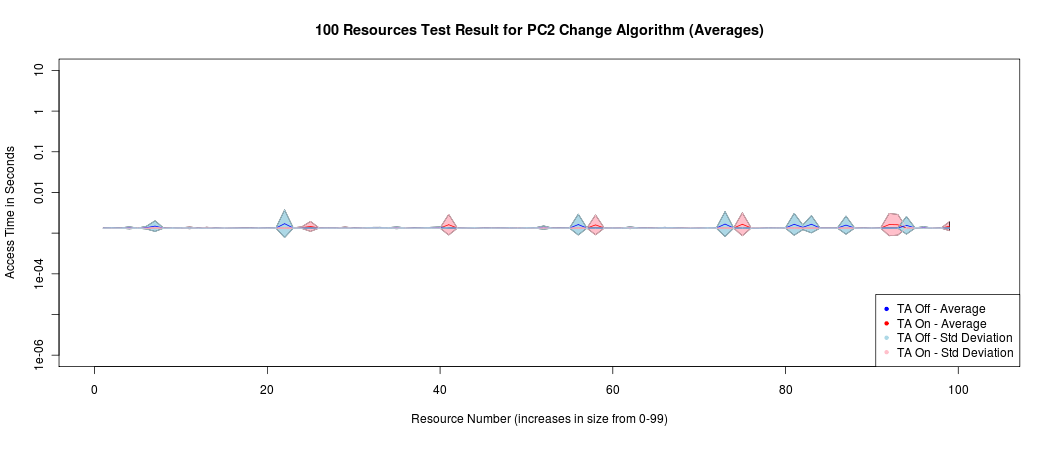}} 
  \end{center}
  \caption{100 resources accessed on PC2. Resource \emph{n} has \emph{n} embedded images.}
  \label{pc200}
\end{figure*}

Figures \ref{ts1} and \ref{ts2} demonstrate the accesses of the 100 resources. The dark blue and red lines indicate the average run time for accessing a resource (in seconds). The filled areas around the lines are the standard deviation ($\sigma$) of the observations over the duration of the experiment.

\subsection{100 Changing Resources: Worst-Case Scenario}
\label{changes}
\begin{figure*}[htp]
  \begin{center}
    \subfigure[Total access time for the 100 static resources on a burdened PC1.]{\label{ts3}\includegraphics[width=150mm]{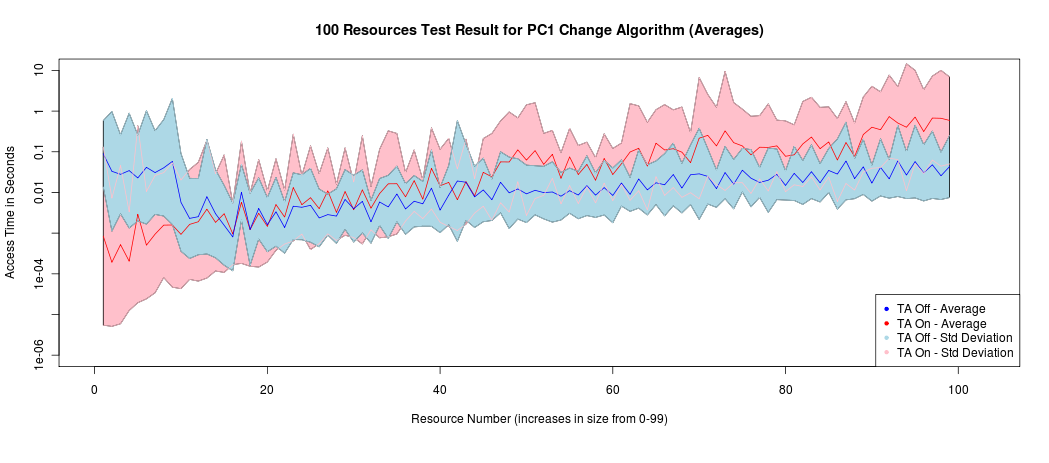}} \\
    \subfigure[Total access time for the 100 changing resources on a burdened PC1.]{\label{tc3}\includegraphics[width=150mm]{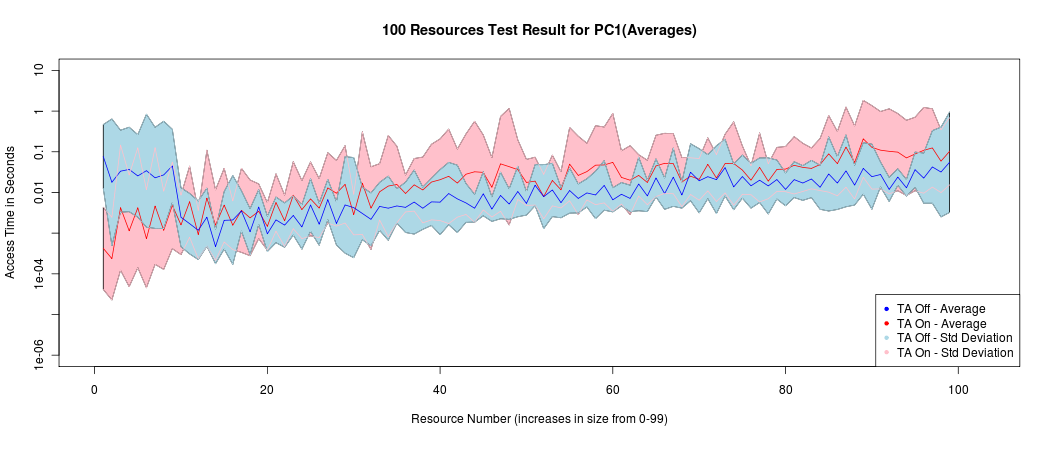}} 
  \end{center}
  \caption{100 resources accessed on a burdened PC1. Resource \emph{n} has \emph{n} embedded images.}
  \label{pc1002}
\end{figure*}

\begin{figure*}[htp]
  \begin{center}
    \subfigure[Total access time for the 100 static resources on a burdened PC2.]{\label{ts4}\includegraphics[width=150mm]{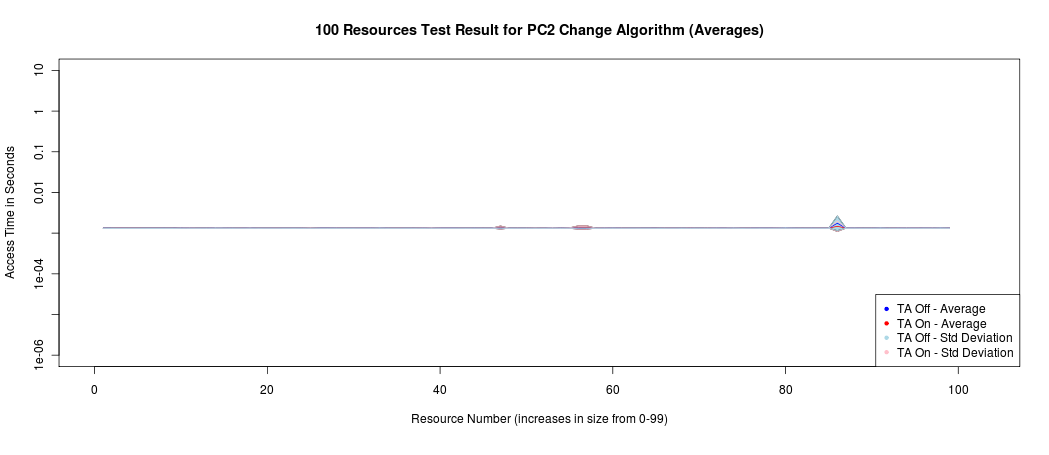}} \\
    \subfigure[Total access time for the 100 changing resources on a burdened PC2.]{\label{tc4}\includegraphics[width=150mm]{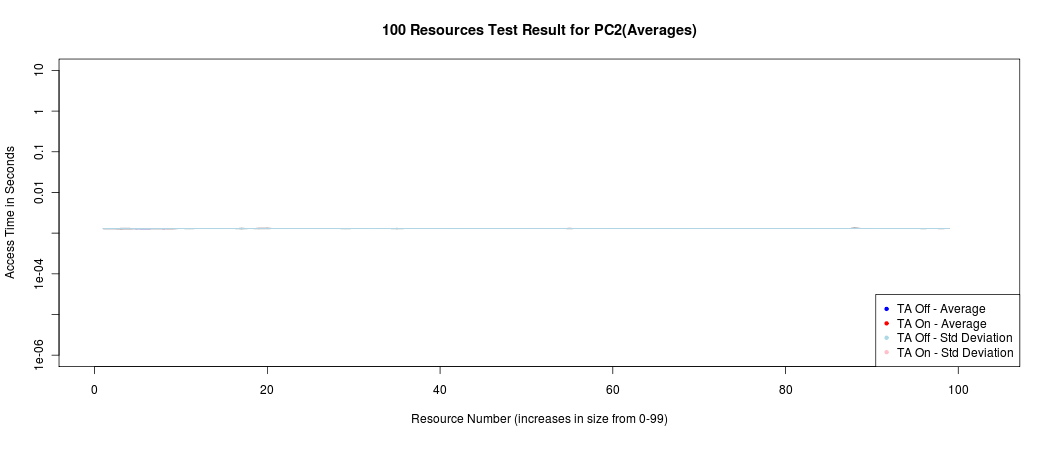}}
  \end{center}
  \caption{100 resources accessed on a burdened PC2. Resource \emph{n} has \emph{n} embedded images.}
  \label{pc2002}
\end{figure*}

The same experiment from Section \ref{static} was run in which each resource changed between runs to provide a ``worst case scenario'' of data connections vs. archiving and run time. A script was executed in between each run in which each resource was updated to make SiteStory archive a new copy of the resource. This means that each access resulted in a new archived copy of each resource. The results of this run are shown in Figures \ref{ts3} and \ref{ts4}. 

Note that Figures \ref{pc1002} and \ref{pc2002} are ``burdened.'' An artificial user load was enduced on the servers to simulate a production environment in which many users are requesting content. A script was run during the test that made curl calls to the server pages to induce the load. The impact of SiteStory operating in a burdened environment is observed in Figures \ref{pc1002} and \ref{pc2002}.

\section{Results}
\label{results}
After the completion of the experiment, the results of each test were analyzed. Immediately, patterns emerge between graphs and tests demonstrating the effect of SiteStory on content server performance. This section explores the results of the tests and makes note of the patterns. From these results, one can conclude whether or not SiteStory affects its host content server in an acceptable manner. 

\subsection{ab Results}
\label{abresults}

For the ApacheBench tests described in Section \ref{ab}, several obvious patterns emerge. Primarily, there is little separation between the total run times of the ab tests when SiteStory is on and when SiteStory is off. One can observe only minor differences in the plotted results. The results differ very little between any given run of the tests, and the averages across the experiment are almost identical in all tests. In the run of 10,000 connections and 1 concurrency, the average total run times were 6.156 seconds when SiteStory was off, and 6.214 seconds when SiteStory was on.  In the run of 10,000 connections and 100 concurrency, the average total run time was 2.4 seconds when SiteStory was off, and 2.42 seconds when SiteStory was on. In the run of 216,000 connections and 1 concurrency, the average run time was 8.905 seconds when SiteStory was off, and 8.955 seconds when SiteStory was on. In the run of 216,000 connections and 100 concurrency, the average total run time was 4.698 seconds when SiteStory was off, and the average total run time was 4.706 when SiteStory was on. This indicates SiteStory does not significantly affect the run time of the ab statistics, and therefore does not affect the performance of the content server with regard to content delivery time. 

Additionally, the concurrency of 1 resulted in more consistent executions across each run whereas the runs with a concurrency of 100 are more inconsistent, as indicated by the spikes in runtime. This could potentially be because of server caching, connection limitations, or even machine memory restrictions. The runs of 100 concurrency also begin with a much longer total run time before dropping significantly and leveling out at runs 9 and 10. 
This is due to additional processes running on the experiment machines that induced extra load in runs 1-8. However, the spikes and inconsistencies do not affect a single run, and do not affect only the runs in which SiteStory is on or those when SiteStory is off. As such, these anomalies are disregarded since they affect both runs.

Finally, the runs of 216,000 connections take much longer to complete than the runs of 10,000 connections -- specifically, 2.736 seconds longer, on average. This is intuitive since more connections should take longer to run. Additionally, the runs of concurrency 1 take 3.9 seconds longer than the runs of 100 concurrent connections. By executing more connections in parallel, the total run time is intuitively shorter. 

The ab test provides evidence that SiteStory does not significantly affect server content delivery time. As such, a production server can implement SiteStory without users observing a noticeable difference in server performance. 

\subsection{100 Resource Results}
\label{staticresults}

The runs of the 100 resources are more interesting, and provide a deeper insight into how SiteStory affects the server's performance than the ab test. This section examines the results of both the static and changing resource tests, as they provide interesting contrasts in performance.The results are listed in Table \ref{table100}.

When comparing the unburdened and burdened results (such as Figure \ref{ts1} vs. Figure \ref{ts3}), it was observed that the average run times are 0.071 and 0.086 seconds higher when the server is under load and SiteStory is off and on, respectively. Additionally, $\sigma$ between the accesses is much greater; 0.1292 and 0.1767 greater when SiteStory is on and off, respectively, as indicated by the wider standard deviation shown on the Figures. 

When comparing the unchanging vs changing resources (such as Figure \ref{ts1} vs. \ref{tc1}), it is apparent that  $\sigma$ is, on average, two times higher for the changing resources than the unchanging resources. (The average $\sigma$ for unchanging resource is 0.0839 and 0.1680 for changing resources.) Additionally, the average access times when SiteStory is off remains approximately the same when the resources change or remain the same. The interesting result is that the average access time increases from 0.15 seconds per GET to 0.21 seconds per GET for the changing resources when SiteStory is on. This is intuitive considering SiteStory needs to re-archive the accessed content during an access when the resource changes. 

When comparing the two machines, PC1 and PC2 (such as Figure \ref{ts1} vs. \ref{tc4}), one sees that PC2 gives a nearly negligible access time, while PC1 gives a measurable access time. This is because PC2 is a dual core machine and can handle the additional load more quickly, while PC1 must context switch between processes, causing an increased delay. 

The most important observation in any of the Figures \ref{ts1} - \ref{tc4} is that the run time of this test is approximately 0.5 seconds higher on average when SiteStory is on vs. when SiteStory is off. This number is reached by comparing the difference in average run time for each test when SiteStory is on vs. off. For each on-off pair, the average difference was taken to reach the approximate 0.5 second difference across all tests. That is, the difference between the average run times of the tests in Figures \ref{ts1} when SiteStory is running (red) vs when SiteStory is off (blue) is 0.08 seconds. When the same comparison is performed across all tests and the average of these results is taken, an overall impact of SiteStory on server performance is realized.

Each Figure begins with SiteStory off taking more time than when SiteStory is on, but this can be attributed to experiment anomaly or similar server access anomaly. Inevitably, the run time when SiteStory is on becomes slower than when SiteStory is off as the resource size increases. This demonstrates that the performance difference of a server when SiteStory is on vs. off is worse when there is a large amount of embedded resources, such as images. PC1's average page access time increases by, on average, 0.006 seconds per embedded image. One could come to the conclusion that servers providing access to image-laden resources would see the biggest performance decrease when utilizing SiteStory. 

\begin{table*}[tp]
\caption{100 Resource Test Results} 
\centering 
\begin{tabular}{c c c c c} 
\hline\hline 
Case & Avg. Unburdened Run Time & Unburdened $\sigma$ & Avg. Burdened Run Time  & Burdened $\sigma$\\ [0.5ex] 
\hline 
\multicolumn{5}{c}{Static Resources}\\
\hline 
PC1, SS Off & 0.121 & 0.0254 & 0.192 & 0.2021 \\
PC1, SS On & 0.206 & 0.1811 & 0.292 & 0.3103 \\
PC2, SS Off & 0.056 & 0.0011 & 0.056 & 0.0001 \\
PC2, SS On & 0.056 & 0.0009 & 0.056 & 0.0001 \\
\hline \hline
\multicolumn{5}{c}{Changing Resources}\\
\hline 
PC1, SS Off & 0.132 & 0.0346 & 0.225 & 0.2174 \\
PC1, SS On & 0.354 & 0.4244 & 0.292 & 0.6137 \\
PC2, SS Off & 0.057 & 0.0021 & 0.056 & 0.0002 \\
PC2, SS On & 0.057 & 0.0016 & 0.056 & 0.0002   \\ [1ex] 
\hline 
\end{tabular}
\label{table100} 
\end{table*}

\section{Conclusions}
\label{conclusion}
In this work,  SiteStory was stress tested and benchmarked. The results of this study have shown that SiteStory does not significantly affect the performance of a server. While different servers and different use cases cause different performance effects when SiteStory is archiving content, the host server is still able to serve sites in a timely manner. The type of resource and resource change rate also affects the server's performance -- resources with many embedded images and frequently changing content are affected most by SiteStory, seeing the biggest reduction in performance. 

These results are observed in Figures \ref{tares} - \ref{pc2002}, as well as Table \ref{table100}. SiteStory does not significantly increase the load on a server or affect its ability to serve content -- the response times seen by users will not be noticeably different in most cases. However, these graphs demonstrate the impact of SiteStory on performance, albeit small -- larger resources with many embedded resources take longer to serve when SiteStory is on as opposed to when SiteStory is off due to the increased processing required of the server. However, the significant finding of this work is that SiteStory will not cripple, or even significantly reduce, a server's ability to provide content to users. Specifically, SiteStory only increases response times by a fraction of a second -- from 0.076 seconds to 0.086 seconds per access when the server is under load, and from 0.15 seconds to 0.21 seconds when the resource has many embedded and changing resources. These increases will not be noticed by human users.


\section{Acknowledgments}
This work is supported in part by NSF grant 1009392 and the Library of Congress. A Corporate Case Study to investigate the feasibility of a transactional archive in a corporate intranet was funded by a Fiscal Year 2011 Innovation Grant from the MITRE Corporation.

MITRE employees Jory T. Morrison and George Despres were integral to the MITRE Innovation Grant and Case Study.

Special thanks to Lyudmila Balakireva, Harihar Shankar, Martin Klein, Robert Sanderson and Herbert Van de Sompel from LANL for the design and development of SiteStory, and their feedback and guidance throughout this experiment.

\bibliographystyle{abbrv}
\bibliography{sitestory_tech_report}

\onecolumn
\appendix 
\section{Sample run of ApacheBench}
\label{abApp}
A sample run of the ApacheBench (ab) is provided below:

\begin{verbatim}
ab -n 1000 -c 1 http://localhost/time.php
This is ApacheBench, Version 2.3 $Revision: 655654 $
Copyright 1996 Adam Twiss, Zeus Technology 
           Ltd, http://www.zeustech.net/
Licensed to The Apache Software Foundation, 
           http://www.Apache.org/

Benchmarking localhost (be patient)
Completed 100 requests
Completed 200 requests
Completed 300 requests
Completed 400 requests
Completed 500 requests
Completed 600 requests
Completed 700 requests
Completed 800 requests
Completed 900 requests
Completed 1000 requests
Finished 1000 requests

Server Software:        Apache/2.2.16
Server Hostname:        localhost
Server Port:            80

Document Path:          /HitPage.html
Document Length:        298 bytes

Concurrency Level:      1
Time taken for tests:   0.260 seconds
Complete requests:      1000
Failed requests:        0
Write errors:           0
Non-2xx responses:      1000
Total transferred:      501000 bytes
HTML transferred:       298000 bytes
Requests per second:    3842.34 [#/sec] (mean)
Time per request:       0.260 [ms] (mean)
Time per request:       0.260 [ms] (mean, across all 
                            concurrent requests)
Transfer rate:          1879.90 [Kbytes/sec] received

Connection Times (ms)
                  min  mean  [+/-sd] median   max 
Connect:        0    0        0.0      0       0
Processing:     0    0        0.0      0       1
Waiting:        0    0        0.0      0       0
Total:          0    0        0.0      0       1

Percentage of the requests served within a certain
                           time (ms)
  50%      0
  66%      0
  75%      0
  80%      0
  90%      0
  95%      0
  98%      0
  99%      0
 100%      1 (longest request)
\end{verbatim}

\end{document}